\documentstyle[multicol,epsf,aps,pra]{revtex}
\begin{document}
\title{Incoherent Energy Transfer within Light-harvesting Complexes}
\author{Julian~Juhi-Lian~Ting 
\footnote{E-mail address:jlting@yahoo.com}
}
\address{
No.38, Lane 93, Sec.2, Leou-Chuan E. Rd., Taichung, 40312 Taiwan, 
Republic of China
}

\date{\today}
\maketitle
\ifpreprintsty
\draft
\else
\tightenlines
\fi
\begin{abstract}
Rate equations are used to model spectroscopic observation
of incoherent energy transfer in light-harvesting 
antenna systems based upon known structures.
A two-parameter two-dimensional model is proposed.
The transfer rates obtained, by matching the fluorescent decay, are 
self-consistent within our model. 
\end{abstract} 

\section{Introduction}

\ifpreprintsty
\else
\begin{multicols}{2}
\fi

We have had a reasonably complete picture of the bacterial light-harvesting (LH)
system recently~\cite{WK,HS}.
Both the inner antenna, LH1, and the outer antenna, LH2, are assembled
from the same modules to form rings. 
Each module consists of two short $\alpha$-helical polypeptides coordinate
one carotenoid and three bacteriochlorophylls (BChls).
The LH2 is composed of 9 units, for 
{\it Rhodopseudomonas acidophila}~\cite{MPFHPCI}, resemble a cylinder,
with an inner diameter $36 \AA$ and an outer diameter $68 \AA$, 
while the LH1  is composed of 16 units,
for {\it Rhodospirillum rubrum}~\cite{KBG}, in order
to accommodate the reaction center (RC). 
The later has an outer diameter $116 \AA$
and a central diameter $68 \AA$. However, the 
exact numbers of both complexes are variable~\cite{WK,KBG,BJMGF}.

Furthermore, the LH2 B850 BChl $a$ form a complete over-lapping ring
in a hydrophobic environment, which reduces the dielectric constant, 
while the B800 BChl $a$ are well separated and are in a polar environment.
When a BChl molecule is excited by light, the energy
can reach equilibrium within about $10 ps$~\cite{TFG}.
A LH2 can function as a storage ring to store the excited
singlet state energy for about $1100 ps$. However, the energy will
transfer to other rings before decaying. The hopping of
energy continues from one ring to another one until a LH1, which contains
the RC, is finally reached. The total trip lasts
for about $ 5$ to $50 ps$~\cite{MPFHPCI,BFGRT,SGBAG}.
Apparently, there is a competition
between energy relaxation and energy transfer.


Historically, relatively few 
physicists have tackled problems of photosynthesis. 
Notably, Montroll used random walk concepts to model 
energy transfer 
amongst antenna rings on a lattice by 
considering its first passage time~\cite{M}.
Later, Hemenger {\it et al.} proposed a more realistic model by
taking inhomogeneous transfer rates and trapping of RCs 
into account~\cite{HPL}.
Interestingly, it is Pearlstein's work which is most often cited
in the literature~\cite{P}. 
In the mean time,
almost all experimentalists try to find some
explanations for their spectral data. 
However, due to lack of precise geometrical
information most efforts are in vain.


Progresses in physics are often made along the line structures - 
energy - dynamics.
A goal of researches nowadays is to find the relation between structural
and spectral information obtained, expecting that 
the function of photosynthesis will be explained 
in terms of its structure, and further
drawing inferences from the model by applying methods of mathematical or 
numerical analysis. 
Recently Timpmann {\it et al.} used a rate equation model to describe
energy trapping and detrapping by the RC~\cite{TZFS}.
However, their antenna has no structure.
Sk{\'a}la {\it et al.} also carried out a series of investigation
by analyzing the spectrum of a more realistic LH1 model~\cite{SK,SJ,DS}.
However, their model is incompatible with the recent structural finding.
In this paper we established a two-parameter model
based on recent structural data. 



\ifpreprintsty
\else
\end{multicols}
\fi
\section{model}

With the known periodical structure, shown
in Fig.\ref{LH1}, we can built, from chemical rate equation,
the following phenomenological model
of energy transfer, 
\begin{eqnarray}
{{d E} \over {d t}} &=& {k^{'}} A_{1} -  ( k^{''} + k_E ) E \; ,  \label{m1}\\
{{d A_1} \over {d t}}&=& k A_{16} - 2 k A_1 + k A_{2} - k^{'} 
A_1 + k^{''} E \; ,\\
{{d A_n} \over {d t}}&=& k A_{n-1} - 2 k A_n + k A_{n+1} \; , 
\qquad n = 2...15 \; , \\
{{d A_{16}} \over {d t}}&=& k A_{15} - 2 k A_{16} + k A_{1} \; ,\label{m4} 
\end{eqnarray}
in which $A_n$s denote the excited BChl dimer, 
$E \equiv P^* B H$ is the excited state,
with B representing the
chlorophyll monomer within the RC, and $P^*$ is the excited special pair of
BChl molecules. 
It is a set of 17 coupled linear differential equations. 
The symmetry of this 
system is broken due to $k^{'} \neq k^{''}$.
A similar model has been proposed by Sk{\'a}la {\it et al.} ~\cite{SJ}.
However, the RC and the antenna 
ring are connected only at one site in the present
model, corresponding to the recent experimental observation.
\ifpreprintsty
\else
\begin{figure}[hbt]
\begin{center}
\epsfxsize=\columnwidth\epsfbox{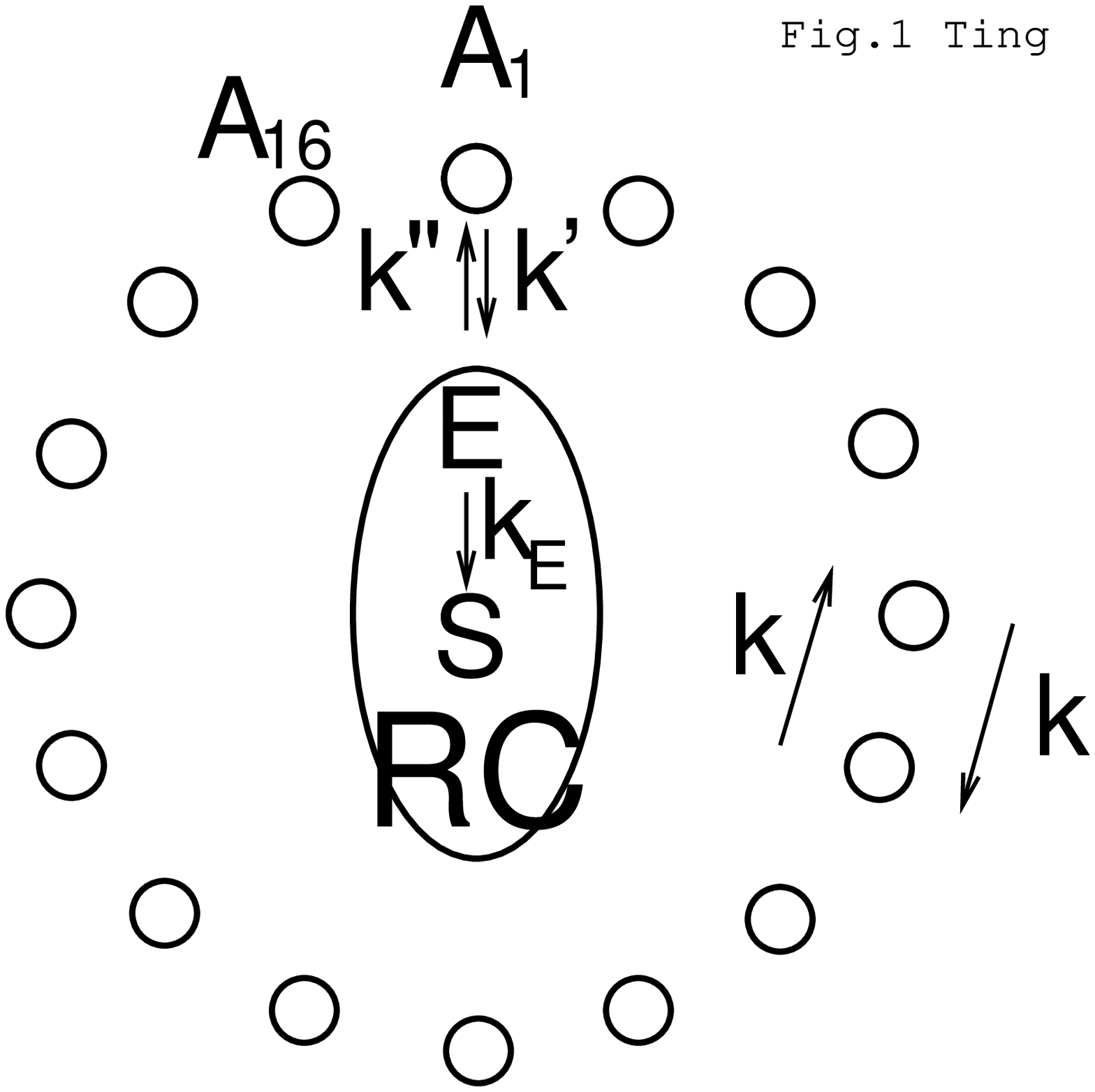}
\caption[1]{Schematic plot of LH 1 and definition of symbols used.}
\label{LH1}
\end{center}
\end{figure}
\fi

In the homogeneous case with the same transition rate amongst the units,
the characteristic polynomial of the above rate-constant-matrix
can always be expressed as
\begin{equation}
P_{16} = P_{16}^1 P_{16}^2 P_{16}^3 P_{16}^4 \;,
\end{equation}
with
\begin{eqnarray}
P_{16}^1 &=& s+2 k \;, \\
P_{16}^2 &=& s^2+4 k s+2 k^2 \;, \\
P_{16}^3 &=& s^4 + 8 k s^3 + 20 k^2 s^2 + 16 k^3 s + 2 k^4 \;, \\
P_{16}^4 &=& s^{10} + ( k_E + k^{''} + k^{'} + 18 k ) s^9 + \nonumber \\
& & ( k^{'} k_E + 18 k k_E + 18 k k^{''} + 
16 k k^{'} + 134 k^2 ) s^8 + \nonumber \\
& & 2 (8 k^{'} k_E + 67 k k_E + 67 k k^{''} + 52 k k^{'} + 266 k^2 ) k s^7 + \nonumber \\
& &2 (52 k^{'} k_E + 266 k k_E + 266 k k^{''} + 
176  k k^{'} + 605 k^2) k^2 s^6 + \nonumber \\
& & 2 ( 176 k^{'} k_E + 605 k k_E +
605 k k^{''} + 330 k k^{'} + 786 k^2 ) k^3 s^5 + \nonumber \\
& & 12 ( 55 k^{'}k_E+ 131 k k_E + 131 k k^{''} + 56 k k^{'}+ 91 k^2) k^4 s^4 + \nonumber \\
& & 4 ( 168 k^{'}k_E + 273 k k_E + 273 k k^{''}+ 84 k k^{'}+ 86 k^2 ) k^5 s^3 
+ \nonumber \\
& &8 ( 42 k^{'}k_E +43 k k_E +43 k k^{''} + 8 k k^{'}+ 4 k^2 ) k^6 s^2 + \nonumber \\
& & 2 ( 32 k^{'}k_E +16 k k_E +16 k k^{''}+ k k^{'} ) k^7 s+2 k^8 k^{'}k_E \; ,
\end{eqnarray}
which is a consequence of the master equation used, and is independent 
of the detail geometrical symmetry. 
The mode controlling the decay to the RC is within $P_{16}^4$, 
since $P_{16}^1$, $P_{16}^2$,
$P_{16}^3$ do not contain $k^{'}$, $k^{''}$ and $k_E$. However, all four parts
will be influenced by the change of $k$.
If one solves this set of differential equations 
by applying the Laplace transformation method,
one finds the solution divides into four distinct groups of decay channels, 
namely, $A_5$-$A_{13}$; 
E-$A_1$-$A_9$; $A_3$-$A_7$-$A_{11}$-$A_{15}$; 
$A_2$-$A_4$-$A_6$-$A_8$-$A_{10}$-$A_{12}$-$A_{14}$-$A_{16}$.
Because the matrix of rate constants is hermitian, all eigenvalues are negative.
Furthermore, no eigenvalues are degenerated, in contrast to 
Sk{\'a}la's model which posses too high degree of symmetry~\cite{DS}.
Letting $k^{'} = k^{''}$ does not results in
additional factorizability although the symmetry of our model is restored.
At $k^{'} = k^{''} = 0$, $P_{16}$ becomes 
\begin{equation}
s {( s + 2 k )}^2 ( s + 4 k ) ( s + k_E )
    ( s^2 + 4 k s + 2 k^2 )^2
    ( s^4 + 8 k s^3 + 20 k^2 s^2 + 16 k^3 s + 2 k^4 )^2 \;. \label{deg}
\end{equation}
It contains 
a zero eigenvalue, which signals the existence of a steady-state solution,
as should be happened without the decay to the RC. Degeneracy of
eigenvalues is introduced as the transition to the RC is decreased.

\section{spectrometry comparison}
\ifpreprintsty
\else
\begin{multicols}{2}
\fi

We can verify our model against experiments: 
The pump-probe spectroscopy measures the difference between two beams, with
\begin{equation}
\Delta D = \Delta \epsilon_A \sum_n A_n + 
\Delta \epsilon_E E\;,    \label{PP}
\end{equation}
being the signal measured. The symbol $\Delta \epsilon$\,s are the 
differences in dielectrical
constants between pump and probe beams of the
corresponding pigments.
By choosing the pump and probe laser frequencies, we can selectively 
detect the population changes of $\sum A_n$ or $E$.
Summing over Eq. (\ref{m1})-(\ref{m4}) 
we know that the decay of the total population should be
$d (\sum A_n ) / dt = - k^{'} A_1 + k^{''} E$.
The measured charge separation rate is
$k_{E} \approx 3.57 \times 10^{11} s^{-1}$ at room temperature, and 
increases by $2$ to $4$
times from $300 K$ to $10 K$ depending on the species chosen~\cite{FMB,MBHMA}.
The ratio of the forward and backward transition to the RC 
is know to be about $25 \%$~\cite{TZFS} for an open  
RC, i.e., the RC BChl dimer (P) is reduced and the
iron quinone electron acceptor is oxidized; $40 \%$ for pre-reduced RC. 
The back-trapping rate can, in principle, be estimated from 
${k^{''} /  k^{'}} =  exp ({- \Delta G / k_B T})$,
with $\Delta G$ the free-energy gap between $A_1$ and $E$ is
estimated from their absorption peaks,  
$k_B$ is the Boltzmann
constant, and $T$ is the absolute temperature.
However, the measured absorption peaks of the excited RC are
broad and imprecise\cite{PF}.
We do not know the absolute values of $k^{'}$ or $k^{''}$ experimentally
since it is difficult to tune 
the laser frequency to distinguish $A_n$ from $E$.
Nor do we know the transition rate between $A_n$s because
transition between the same species cannot
be measured directly.  
Furthermore, at room temperature, energy equilibrium within the
antenna interferes with the trapping process.
Therefore we have taken $k$ and $k^{'}$ as 
parameters and fit the slow mode of fluorescence decay of excited population 
observed, i.e. $200 ps$~\cite{TFG,BFGRT,SGBAG}.
Thus, the absolute value of the largest eigenvalue 
should be about $3 / 200 ps = 1.5 \times 10^{10} s^{-1}$.
A computer code is written to scan all combinations of $k$ and $k^{'}$,
with $k^{''} = k^{'} / 5$, for the largest eigenvalue to be smaller than
$ - 1.5 \times 10^{10} s^{-1}$ 
between $ - 10^{8} s^{-1}$ to $ - 10^{15} s^{-1}$. Interestingly, we
find all possibility happened at $ k = k^{'}$ and $ k > 6.97 \times
10^{11} s^{-1}$ 
for $k^{''} = k^{'} / 5$. 
Presumably, it is an extremum of $P_{16}^4$.
At the lowest $k$, we can match the required 
$200 ps$ decay whose
curve is plotted at Fig.\ref{LH2}.
If $k^{''} = k^{'} / 4$, we obtained $k = 7.25 \times 10^{11} s^{-1}$.
That $k$ has to be equal to $k^{'}$ 
might sound peculiar in view of the geometrical distance
between $A_1$ and $RC$ is less than the distance between $RC$ and other
$A_n$s \cite{KBG}.
However, the species for donor and acceptors 
are different at these two cases. 
There are possibilities that the final hopping rate are still the same.

\ifpreprintsty
\else
\begin{figure}[hbt]
\begin{center}
\begin{minipage}[hbt]{8.5cm}
\epsfxsize=\columnwidth\epsfbox{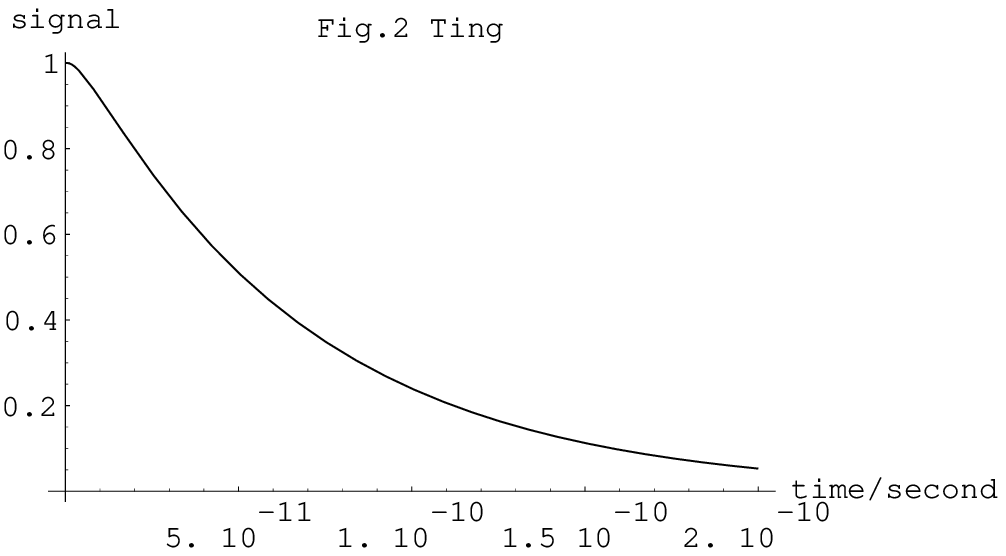}
\caption{Numerically calculated pump-probe 
signal from Eq. (\protect{\ref{PP}})
at $k = 6.97 \times 10^{11}$, 
$k_E = 3.57 \times 10^{11}$, $k^{'} = k$, $k^{''} = k^{'}/5$,
$\Delta \epsilon_A = 1$, 
$\Delta \epsilon_E = 0$. 
The initial condition is $A_5 = 0.2$,
$A_7 = 0.4$, $A_{10} = 0.3$, $A_{12} = 0.1$, while other sites are not excited
at $t = 0$.}
\label{LH2}
\end{minipage}
\end{center}
\end{figure}
\fi

The transfer of excitation energy requires 
coupling between the emitting molecule and the ground state molecule.
At an intermolecular separation involved between 
$10 \AA$ to $ 100 \AA$, long-range resonance
transfer of electronic excitation arises from coupling between the transition 
dipoles  of the donor and  the acceptor, which is the 
F{\"o}ster theory~\cite{F,W}.
Since the BChl $Q_y$ dipoles lie in the same plane, we have
\begin{equation}
k ( R ) \propto {1 \over {\tau_F}} ( {R_0 \over R })^6 \; ,
\end{equation}
in which  $R_0$,
measures transfer efficiency, is the F{\"o}ster radius. 
van Grondelle gave $R_0 = 90 \AA$ for 
the BChl 875 to BChl 875 energy transfer 
and a fluorescence life
time, $\tau_F$, about $3000 ps$ or 
slightly higher~\cite{RVG,CSJ}.
If a putative 
separation distance between interacting BChl $a$ dimers 
$\approx 17.5 \AA$
is used~\cite{KBG} we obtain
an estimation of $k \approx 6.17 \times 10^{12} s^{-1}$. This number is
about an order of magnitude higher than the value obtained from our model.
However, the pairwise energy transfer is about $ 1 ps$ according to our 
calculation~\cite{SGBAG}.
On the other hand, from the value of $k$ obtained here, by
fitting the $200 ps$ decay as well as the $\tau_F$,
we estimated the F{\"o}ster radius to be
$26.8 \AA$. This result is consistent within our model 
since we assume only nearest neighbour transition.
Further, since we put the population at 
the antenna at $t = 0$ for our calculation,the rising time is 
infinitely short,
instead of having some instrumental limits as observed experimentally.
Although the light wave length is much larger than the ring size, the ring
still might receive energy in localized form by energy transfer from 
other rings as the initial condition we used in Fig.\ref{LH2}.
Table I provides a list of all eigenvalues and corresponding amplitudes 
obtained from our model.
From the table, we found that the largest eigenvalue mode
is important, not only for its large separation from the other
eigenvalues but also for its corresponding large amplitude.

\ifpreprintsty
\else
\begin{table}[bht]
\begin{center}
\begin{minipage}[bht]{7cm}
\begin{tabular}{ l l }
amplitude & eigenvalue \\
\hline
$ -0.0070   $& $-2.9707 \times 10^{12}$ \\
$ -0.0006   $& $-2.7443 \times 10^{12}$ \\  
$ ~~0.0000   $& $-2.6819 \times 10^{12}$ \\
$ ~~0.0221   $& $-2.4614 \times 10^{12}$ \\  
$ ~~0.0000   $& $-2.3797 \times 10^{12}$ \\
$ -0.0200   $& $-2.0143 \times 10^{12}$ \\  
$ ~~0.0000   $& $-1.9275 \times 10^{12}$ \\
$ ~~0.0124   $& $-1.4844 \times 10^{12}$ \\  
$ ~~0.0000   $& $-1.3940 \times 10^{12}$ \\
$ -0.0086   $& $-9.5864 \times 10^{11}$ \\
$ ~~0.0000   $& $-8.6054 \times 10^{11}$ \\
$ -0.0159   $& $-5.5606 \times 10^{11}$  \\
$ ~~0.0000   $& $-4.0829 \times 10^{11}$  \\
$ ~~0.0034   $& $-3.8570 \times 10^{11}$ \\  
$ -0.0716   $& $-1.4875 \times 10^{11}$ \\
$ ~~0.0000   $& $-1.0611 \times 10^{11}$ \\  
$ ~~1.0858   $& $-1.5107 \times 10^{10}$ 
\label{tab}
\end{tabular}
\caption{Eigenvalues for LH 1 for Fig.\protect{\ref{LH2}}.
In a time-resolved experiment the relaxation rates correspond to the 
lifetimes observed from antenna fluorescence or bleaching kinetics.
}
\end{minipage}
\end{center}
\end{table}
\fi

We have also introduced inhomogeneity, due to geometrical distortion, into the
rate constant. However, even at large
distortion, the basic character of the spectrum is not altered considerably.
If the criteria for $k = k^{'}$ can be established, we can further reduce
the free parameters in our model.

\section{Conclusion}



In summary, a physicist's approach \cite{GLNV} of incoherent energy transfer
within an antenna ring is taken by considering 
a two-parameter two-dimensional model.
This model differs from the one presented by Sk{\'a}la {\it et al.}.
The reality might be somewhere between these two models.
In our model, we numerically found $k$ has to be equal to $k^{'}$.
Furthermore, we are able to calculate analytically some of the eigenvalues 
and demonstrate explicitly that there is a
mode for decaying to the RC and other three modes.
However, this result of mode separation
depends upon the exact number of unit involved in the
ring. Therefore should not be important. 
Perhaps we should interpret the finding
as: $P_{16}^1$, $P_{16}^2$, $P_{16}^3$ are redundant, 
since $P_{16}^4$ contains $k_E$ which should be important. 
A ring of $16$ units is huge. 
The only purpose for such a large antenna is to accommodate the $RC$.

Finally we remark that it is possible to extend a two-dimensional
random walk model of energy transfer into a quasi-three-dimensional one,
in view of the recent structural finding, with a recent result of random walk
on bundled structures by Cassi and Regina\cite{CR2}.
Furthermore, this theoretical result should be able to be verified 
experimentally using its spectral dimension by measurements
involving diffusion processes such as time-resolved spectroscopy of
nearest-neighbours energy transfer.
Other light-harvesting models  and mechanisms are under further investigation.



\ifpreprintsty
\begin{figure}
\caption[1]{Schematic plot of LH 1 and definition of symbols.}
\label{LH1}
\end{figure}
\begin{figure}
\caption[1]{Numerical calculated pump-probe 
signal from Eq. (\protect{\ref{PP}})
at $k = 6.97 \times 10^{11}$, 
$k_E = 3.57 \times 10^{11}$, $k^{'} = k$, $k^{''} = k^{'}/5$,
$\Delta \epsilon_A = 1$, 
$\Delta \epsilon_E = 0$. 
The initial condition is $A_5 = 0.2$,
$A_7 = 0.4$, $A_{10} = 0.3$, $A_{12} = 0.1$, while other sites are not excited
at $t = 0$.}
\label{LH2}
\end{figure}

\begin{table}[p]
\begin{center}
\begin{minipage}[h]{6cm}
\begin{tabular}{ l l }
amplitude & eigenvalue \\
\hline
$ -0.0070   $& $-2.9707 \times 10^{12}$ \\
$ -0.0006   $& $-2.7443 \times 10^{12}$ \\  
$ ~~0.0000   $& $-2.6819 \times 10^{12}$ \\
$  ~~0.0221   $& $-2.4614 \times 10^{12}$ \\  
$  ~~0.0000   $& $-2.3797 \times 10^{12}$ \\
$ -0.0200   $& $-2.0143 \times 10^{12}$ \\  
$  ~~0.0000   $& $-1.9275 \times 10^{12}$ \\
$  ~~0.0124   $& $-1.4844 \times 10^{12}$ \\  
$  ~~0.0000   $& $-1.3940 \times 10^{12}$ \\
$ -0.0086   $& $-9.5864 \times 10^{11}$ \\
$  ~~0.0000   $& $-8.6054 \times 10^{11}$ \\
$ -0.0159   $& $-5.5606 \times 10^{11}$  \\
$  ~~0.0000   $& $-4.0829 \times 10^{11}$  \\
$  ~~0.0034   $& $-3.8570 \times 10^{11}$ \\  
$ -0.0716   $& $-1.4875 \times 10^{11}$ \\
$  ~~0.0000   $& $-1.0611 \times 10^{11}$ \\  
$  ~~1.0858   $& $-1.5107 \times 10^{10}$ 
\label{tab}
\end{tabular}
\caption{Eigenvalues for LH 1 for Fig.\protect{\ref{LH2}}.
In a time-resolved experiment the relaxation rates correspond to the 
lifetimes observed from antenna fluorescence or bleaching kinetics.
}
\end{minipage}
\end{center}
\end{table}


\begin{figure}[p]
\begin{center}
\epsfxsize=\columnwidth\epsfbox{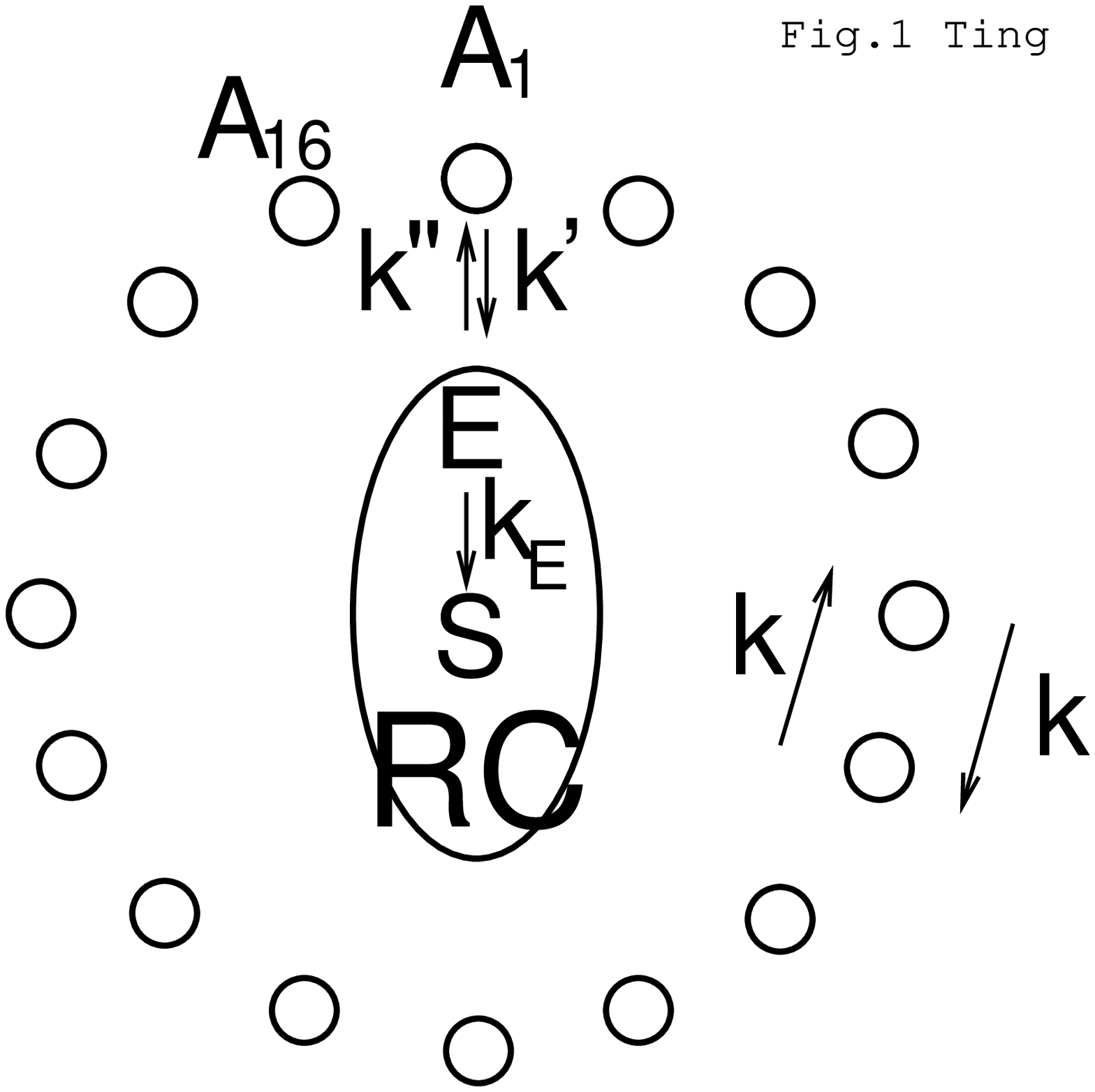}
\end{center}
\end{figure}
\newpage
\begin{figure}[p]
\begin{center}
\epsfxsize=\columnwidth\epsfbox{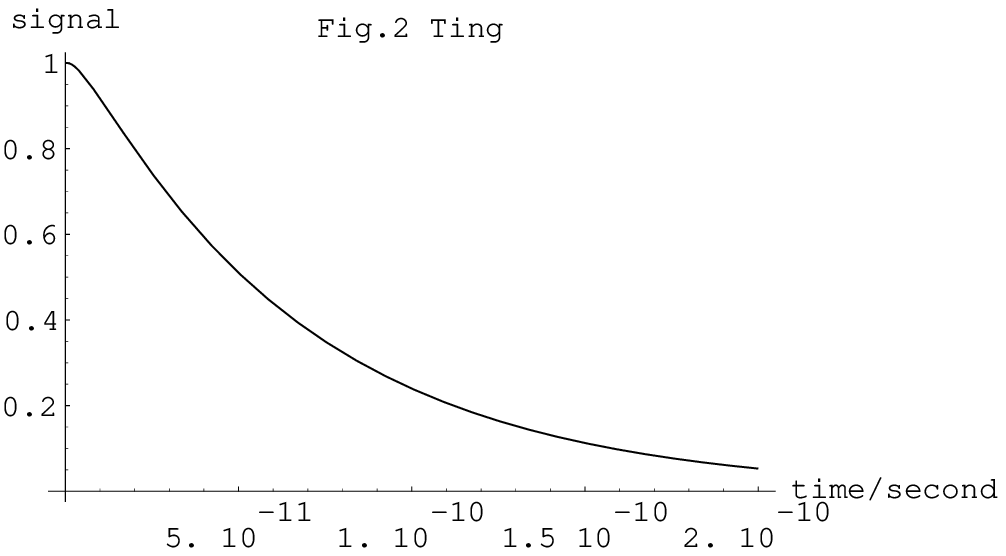}
\end{center}
\end{figure}
\else
\end{multicols}
\fi

\end{document}